\documentclass[preprint]{vgtc}               % preprint
\ifpdf%                                % if we use pdflatex
  \pdfoutput=1\relax                   % create PDFs from pdfLaTeX
  \usepackage{graphicx}                % allow us to embed graphics files
  \DeclareGraphicsExtensions{.pdf,.png,.jpg,.jpeg} % for pdflatex we expect .pdf, .png, or .jpg files
\else%                                 % else we use pure latex
  \usepackage{graphicx}                % allow us to embed graphics files
  \DeclareGraphicsExtensions{.eps}     % for pure latex we expect eps files
\fi%

\graphicspath{{figures/}{pictures/}{images/}{./}} % where to search for the images
\usepackage{savesym}
\usepackage{amsfonts}
\usepackage{amssymb}
\usepackage{soul}
\usepackage{amsmath}
\usepackage{subcaption}
\savesymbol{subfigureautorefname}
\DeclareMathOperator*{\argmax}{argmax}
\DeclareMathOperator*{\argmin}{argmin}
\usepackage{makecell}
\usepackage{microtype}    
\PassOptionsToPackage{warn}{textcomp}  % to address font issues with \textrightarrow
\usepackage{textcomp}% use better special symbols
\usepackage{mathptmx}                  % use matching math font
\usepackage{times}                     % we use Times as the main font
\usepackage{cite}                      % needed to automatically sort the references
\usepackage{tabu}                      % only used for the table example
\usepackage{booktabs}   
\usepackage{color}
\definecolor{mattblue}{RGB}{23,55,180}

\onlineid{0}

\vgtccategory{Research}

\vgtcinsertpkg

\preprinttext{To appear in an IEEE VGTC sponsored conference.}

\title{Visually Analyzing and Steering Zero Shot Learning}

\author{Saroj Sahoo\thanks{e-mail: saroj.k.sahoo@vanderbilt.edu}\\ %
        \scriptsize Vanderbilt University %
\and Matthew Berger\thanks{e-mail: matthew.berger@vanderbilt.edu}\\ %
     \scriptsize Vanderbilt University %
     }

\teaser{
\centering
  \includegraphics[width=.8\linewidth]{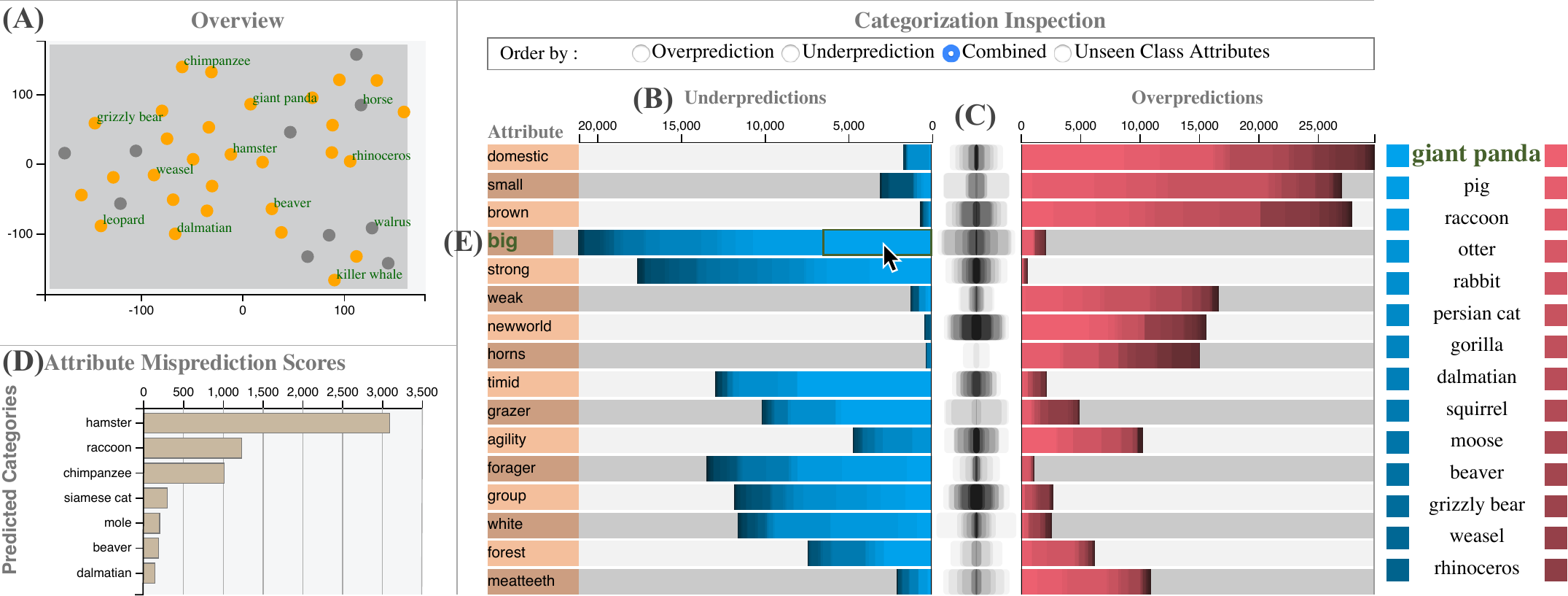}
  \caption{Our visual interface for diagnosing and steering the procedure of zero-shot learning. A scatterplot overview (A) enables the user to select categories for more detailed inspection. The main view (B-C) encodes category predictions with respect to attributes -- human-interpretable properties that describe categories. For a given category and attribute, the user can inspect details (D) on category mispredictions made by the model. Based on these views, the user can modify attribute importance (E) to steer the model and mitigate errors when applied to data associated with categories not seen during training.}
  \label{fig:teaser}
}

\abstract{We propose a visual analytics system to help a user analyze and steer zero-shot learning models. Zero-shot learning has emerged as a viable scenario for categorizing data that consists of no labeled examples, and thus a promising approach to minimize data annotation from humans. However, it is challenging to understand where zero-shot learning fails, the cause of such failures, and how a user can modify the model to prevent such failures. Our visualization system is designed to help users diagnose and understand mispredictions in such models, so that they may gain insight on the behavior of a model when applied to data associated with categories not seen during training. Through usage scenarios, we highlight how our system can help a user improve performance in zero-shot learning.%
} % end of abstract

\CCScatlist{
  \CCScatTwelve{Zero-shot learning}{Visu\-al\-iza\-tion}{Visu\-al\ Analytics}{Model Steering}{}
}

\begin{document}

\firstsection{Introduction}

\maketitle

% What is ZSL, what are its challenges
Many of the recent successes in machine learning are owed to the ample amounts of supervised data provided by humans. Yet humans cannot simply be treated as an unlimited resource of supervision, as it is costly, in terms of both time and cognitive load, for humans to annotate data. Hence, the problem of zero-shot learning (ZSL)~\cite{xian2018zero} has emerged as one way to support scenarios in machine learning where human supervision is scarce. Specifically, the typical setup in ZSL is to build a model on a dataset where categories are provided (\emph{seen}), in order to apply the model to data associated with categories not provided during the model's construction (\emph{unseen}). To solve this problem, ZSL models typically rely on \emph{attributes} -- semantic and human-nameable properties that characterize categories -- to establish a relationship between categories associated with data seen during training, and unseen categories. However, this transfer between seen and unseen categories faces several challenges. Some attributes might not be particularly informative for the learning task at hand, while relevant attributes might be difficult to accurately model. A deeper understanding of attributes can help support the analytic tasks of both consumers and developers of ZSL methods, in terms of analyzing where and why mispredictions occur, and how to edit the model to mitigate error when applied to unseen categories.

% Our work: who are we addressing, distinctions between typical VA for ML, and what we are addressing
In this work, we propose a visual analytics system to analyze and steer ZSL models. Our approach is attribute-centric: we diagnose mispredictions in terms of attributes to convey potential failure modes of ZSL. Although significant work has been recently developed in the visual analytics community for diagnosing and understanding the performance of machine learning models~\cite{ren2016squares,krause2017workflow}, ZSL presents a unique challenge. Specifically, in the visualization, we do not have \emph{any} access to data associated with categories for which the model will ultimately be used. Adhering to the traditional ZSL setup, our visualization only has access to data associated with seen categories. On the other hand, we also assume access to category-level attributes, both for seen and unseen categories. The main goal of our work is to use this provided information to help the user -- be it model developer or consumer -- make good decisions on how to \emph{steer} the model, through analyzing, identifying, and modifying attributes in terms of their \emph{reliability} for categorization.

The main contributions of our work can be summarized as follows: (1) We show how to extract information from predictions made by a ZSL model that help explain model errors. (2) We summarize this information in a visualization design that supports the identification of errors in seen categories, and how such errors might transfer to data associated with unseen categories. (3) Our design supports model steering, and we show how a modification of ZSL can support such user feedback. (4) We show through usage scenarios how our system can support a user in understanding a ZSL model for image categorization, and how to improve its performance.
% \begin{itemize}
% 	\item We show how to extract information from predictions made by a ZSL model that helps explain errors in the model.
% 	\item We summarize this information in a visualization design that can help users quickly understand causes of error for seen categories, and how such errors might transfer to data associated with unseen categories.
% 	\item Our visualization design supports the user in steering the model, and we show how a modification of ZSL can support such user feedback.
% 	\item We show through usage scenarios how our system can support a user in understanding a ZSL model for image categorization, and how to improve its performance.
% \end{itemize}

\section{Related Work}

% First: VA for understanding ML models
Our approach is related to methods in visual analytics for understanding, diagnosing, and steering machine learning models; please see Hohman et al.~\cite{hohman2019gamut} for an overview on the types of questions users have when interacting with models. Within model understanding, approaches based on \emph{local interpretability} help explain why a prediction was made through locally-built model proxies~\cite{ribeiro2016should,ribeiro2018anchors}, using a learned rule list to explain predictions~\cite{ming2018rulematrix}, diagnosing why an error occurred in a prediction~\cite{krause2017workflow,chen2018anchorviz}, or potential changes to features that impact a prediction~\cite{krause2016interacting}. On the other hand, \emph{global interpretability} focuses on summarizing a set of predictions. Prior work has considered how to present the confidence of classifiers in multi-class settings~\cite{ren2016squares}, grouping feature subsets into coherent predictions~\cite{krause2017workflow}, and visually analyzing discriminative features~\cite{krause2014infuse,zhang2018manifold}.

% Second: VA for steering ML models
Visual analytics for model steering focuses on how to take approaches that aid in understanding models to inform the user on how to edit the model to improve performance. Approaches in model steering must identify information, from both the model and data, that is useful for users in editing a model~\cite{tam2016analysis}. Prior work has considered this for constructing decision trees~\cite{van2011baobabview}, while other work has developed methods for building ensemble models through visualizing confusion matrices~\cite{talbot2009ensemblematrix} and joint scatterplots of data and models~\cite{schneider2017integration}. Our steering scheme is inspired by work that learns distance functions via direct manipulation~\cite{brown2012dis,hu2013semantics}.

% A paragraph on how existing approaches are not immediately applicable to our scenario
Existing approaches for visually understanding and steering models, however, are not directly applicable to the case of ZSL, for several reasons. First, all such approaches assume that labeled data samples exist for all categories apriori. In ZSL, we need to determine what information from the training data, and resulting model, is useful to present to a user, without access to data of unseen categories. This is necessary for making effective decisions on adjusting the model, in order to improve model performance when eventually deployed. Secondly, techniques that aim to understand predictions via feature importance, either at a local~\cite{ribeiro2016should,krause2017workflow} or global~\cite{krause2014infuse} level, again due to lack of data for test categories. However, we assume access to attributes of unseen categories, and thus we utilize the attribute vectors of categories, in addition to the data mapping into the attribute space, for diagnosing errors in predictions.

\section{Visual Analytics for Zero-Shot Learning}

In this section we describe the problem of zero-shot learning (ZSL), and the tasks we aim to address in our visualization design. ZSL is applicable to numerous problem domains~\cite{wu2014zero,qiao2016less}, but to focus our work, we consider multi-class image categorization, which has received significant attention~\cite{xian2018zero}. In this setting, the goal is to train a model on images of \emph{seen} categories, that can recognize images associated with categories \emph{unseen} at training. The transfer of knowledge from seen to unseen categories is done with the help of attributes -- a set of human-nameable properties that describes a category. For instance, if our categories consist of animals, then attributes are properties of animals, e.g. \textit{furry}, \textit{paws}, \textit{fast}, that are shared between animals, e.g. \textsf{bobcat} and \textsf{siamese cat} both have \textbf{paws}, but the latter is \textit{faster} than the former. Attributes enable a common space between seen and unseen categories that can be used for categorization: given an image, we predict its attributes, and then find the category whose attributes are most similar to the image's attributes. Given this setting, a major goal of ZSL is the characterization of unseen categories as \textit{combinations of attributes} from seen categories. For instance, suppose category \textsf{leopard} was unseen, but \textsf{bobcat} and \textsf{siamese cat} were seen. All three categories have many attributes in common, e.g. paws, quadrupedal, that would allow the model to distinguish an image of a \textsf{bobcat} from a different, unseen category, e.g. a \textsf{whale}.

However, the performance of ZSL is highly dependent on its ability to accurately model attributes. In the above example, it is possible for a model to underpredict the attribute \textit{fast} for unseen category \textsf{leopard}, and thus confuse it as a different, but related, category e.g. \textsf{persian cat}. These issues are well-understood in the community~\cite{lazaridou2015hubness,paul2019semantically}. Namely, ZSL methods can suffer from the \textit{hubness} problem, where an image's predicted attributes become a hub for category attributes, reducing the model's discriminative power, while there is also a tendency to \textit{bias} the image's attributes towards the seen categories and thus reduce the effectiveness in mapping to unseen categories. Yet, there remains a lack of methods for more detailed inspection, that can help diagnose problems in ZSL. In particular, we would like to gain insight on where a model is likely to \textit{overpredict} or \textit{underpredict} certain attributes. This can be useful to understand how such a model might behave on unseen categories, and in particular, what can be done to modify the model to prevent potential errors. 

These issues inform the goals we aim to address:\\
\textbf{G1} -- Understand a ZSL model on seen categories, namely incorrect classifications, and the relationship to attribute predictions.\\
\textbf{G2} -- Depict model behavior for unseen categories, e.g. how the model might make poor attribute predictions.\\
\textbf{G3} -- Enable model steering: prioritize attributes for prediction in unseen categories.

We address these goals by supporting the following tasks:\\
\textbf{T1} -- Provide category overviews, and category selection, for more detailed, downstream analysis of categories [G1,G2].\\
\textbf{T2} -- Enable attribute-centric exploration to highlight errors in seen categories [G1].\\
\textbf{T3} -- Compare attribute-based errors in seen categories with unseen categories [G2].\\
\textbf{T4} -- Support user editing of attribute importance to mitigate prediction error in unseen categories [G3].

\section{Visualization Design}

In this section we discuss the ZSL model we use, the data collected from the model, as well as the design of the visualization.

\subsection{ZSL Model and Data}
\label{sec:zsl-model-data}

Our visualization is designed to support ZSL models that learn to transform the input data into an attribute space. To this end, our model is based on Akata et al.~\cite{akata2015evaluation}. This approach can be broken down into two main components: learning a mapping from input to attribute space, and optimizing a max-margin loss driven by an attribute-based compatibility function.

\textbf{Mapping to Attribute Space.} In this step, our goal is to define a function $f$ that maps the $d$-dimensional input $\mathbf{x}$ to the $a$-dimensional attribute space. We adopt Akata et al.~\cite{akata2015evaluation} and represent $f$ as a 2-layer neural network, though our design could support other, related, approaches~\cite{zhang2015zero,akata2015evaluation,romera2015embarrassingly,lampert2013attribute}.

\textbf{Optimizing a Compatibility Function.} Given the mapping, in this step our goal is to optimize a categorization criterion. This requires a way to define a compatibility between an input data instance, and a category -- both represented as attributes. The compatibility function, $s$, we use is the dot product between points in the attribute space, specifically : $s(\mathbf{z}_i,\mathbf{z}_j) = \mathbf{z}_i^{\intercal}\mathbf{z}_j$,
where $\mathbf{z}_i$ and $\mathbf{z}_j$ are $a$-dimensional attribute vectors. Combined with the attribute mapping function, this permits us to measure the similarity between data $\mathbf{x}$ and an attribute vector for a category $\mathbf{z}$ via $s(f(\mathbf{x}),\mathbf{z})$. Given $s$, the loss we use to optimize for $f$ maximizes the margin~\cite{akata2015evaluation} of compatibilities between an input's ground truth category denoted $y_i$, and all other seen categories denoted $y \in S$:
\begin{equation}
\argmin_{f} \sum_{i=1}^n \max_{y \ne y_i} \lfloor s(f(\mathbf{x}_i),\mathbf{z}_{y}) - s(f(\mathbf{x}_i),\mathbf{z}_{y_i}) + \eta \rfloor_{+},
\label{eq:maxmargin}
\end{equation}
where $\eta$ is a margin hyperparameter and $\lfloor x \rfloor_{+}$ returns $x$ if the expression is positive, and zero otherwise. Once optimized, we may use both the learned mapping $f$ and the compatibility function $s$ to categorize a data input $\mathbf{x}$ from a set of unseen categories, denoted $U$, via 
$\argmax_{y \in U} s(f(\mathbf{x}),\mathbf{z}_y)$.

In order to visually diagnose prediction errors that are made by the model, we first take a closer look at the max-margin loss for a single data instance $\mathbf{x}_i$ (discarding the margin, as it does not affect predictions):
\begin{equation}
    \max_{y \ne y_i} s(f(\mathbf{x}_i),\mathbf{z}_{y}) - s(f(\mathbf{x}_i),\mathbf{z}_{y_i}) = \max_{y \ne y_i} f(\mathbf{x}_i)^{\intercal} (\mathbf{z}_y - \mathbf{z}_{y_i}).
\end{equation}
Assume that the model predicts some category $y$ that renders the expression positive, indicative of a misprediction. As we are using a dot product-based similarity, the above expression can be rewritten as:
\begin{equation}
\sum_{k=1}^a p^k_y(\mathbf{x}_i),
\end{equation}
where the summand is $p^{k}_{y}(\mathbf{x}_i) = f_k(\mathbf{x}_i) (\mathbf{z}_{y,k} - \mathbf{z}_{y_i,k})$ for each attribute $k$. Whenever $p^{k}_{y}(\mathbf{x}_i) > 0$, this indicates the mapping $f$ is poor -- we are either \emph{overpredicting} or \emph{underpredicting} this attribute for the given data $\mathbf{x}_i$, relative to its incorrect prediction $y_i$. Specifically, we are overpredicting if $f(\mathbf{x}_i) > 0$, and underpredicting if $f(\mathbf{x}_i) < 0$. It is precisely these summands we would like the user to inspect in our visual interface, to understand mispredictions, e.g. what attributes are being overpredicted/underpredicted, and by how much (\textbf{T2}).

%% Can add the following in supplement materials as Training details or something?

% The scores $p$ are most relevant when applied to data of unseen categories. Hence, constructing a ZSL model just on the training data will not leave us with any particularly useful information, as we do not have access to data from unseen categories. To address this, we reorganize the training dataset into 3 folds, where in each fold we partition the data by categories that are used to train the model, and treat the remaining categories, and their corresponding data, as unseen. We then train the model on data of seen categories, followed by computing and storing values $p^{k}_{y}(\mathbf{x})$ whenever \textit{both} data $\mathbf{x}$ of an unseen category $y$ (for the given fold) is incorrectly predicted, and attribute $k$ is contributing towards this misprediction. In addition, we also save the attributes of all unseen categories in $U$.

\subsection{Category Overview}

% The data that we collect from our model, $p^{k}_{y}(\mathbf{x})$, comprises all images $\mathbf{x}$, seen categories $y \in S$, and attributes $k$, in addition to the unseen category attributes. As this is a large and complex dataset, a visual interface that enables the user to explore the data at multiple levels of detail is essential. To this end,

We first present to the user an overview with respect to all categories (\textbf{T1}). Specifically, we perform t-SNE~\cite{van2009learning} with respect to the category attribute vectors, obtaining a 2D scatterplot of categories -- both seen and unseen. We use t-SNE to ensure that local structure is retained in the projection, namely, categories with similar attributes will be close to one another. In the scatter plot we visually encode the seen categories with the color \textit{blue} whereas we encode unseen categories with color \textit{red}. We enable selection of seen categories via rectangular brushing, and unseen categories via clicking on individual points, where upon selection, seen categories are visually encoded as \textit{orange} and unseen categories as \textit{grey}, please see Fig.~\ref{fig:teaser}(A).

\begin{figure}[t]
\centering
\includegraphics[width=0.6\linewidth]{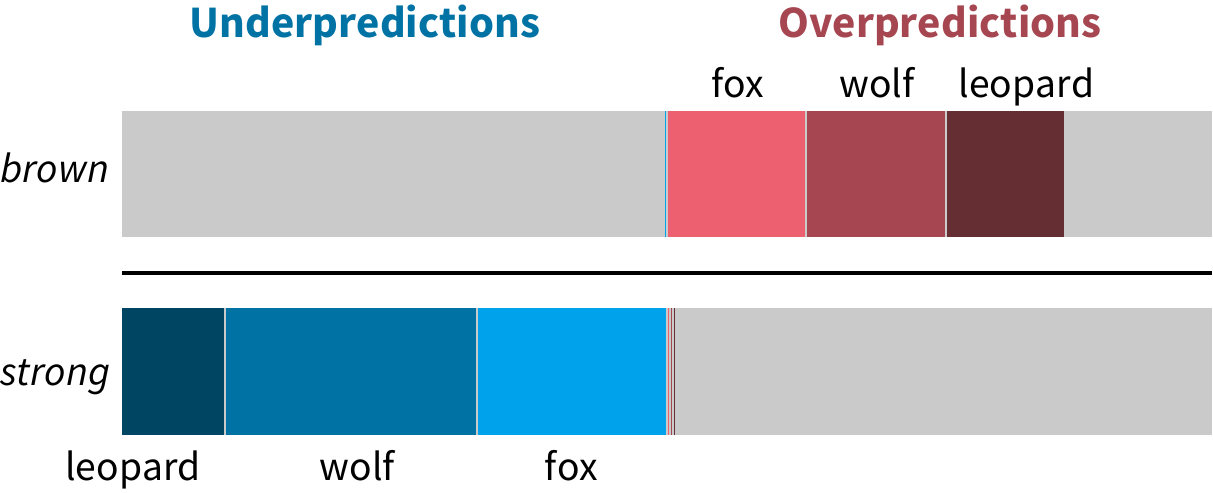}
\caption{We visually encode underpredictions and overpredictions of attributes via a diverging and stacked bar plot, where underpredictions of categories are stacked to the left, and overpredictions to the right, and each row encodes an attribute.} 
\label{fig:main}
\end{figure}

\subsection{Visually Exploring Attributes of Seen Categories}
\label{seen_cat}

Our main view summarizes mispredictions made by the model, captured by the scores $p^{k}_{y}(\mathbf{x})$. This view is prioritized by attributes, so that the user can explore incorrect predictions due to attributes with respect to user-selected categories. In particular, we would like to distinguish overpredictions from underpredictions. To this end, for a given attribute $k$ and category $y$, we sum up all scores for data whose categories are $y$, individually for over/underpredictions:
\begin{equation}
    q^{+/-}_{k,y} = \sum_{(\mathbf{x}_i,y_i) \in D_s} p^{k}_{y}(\mathbf{x}) \quad  
    \begin{cases}    
        \text{if} \quad y_i = y \; , \; f(\mathbf{x}_i) > 0\\
        \text{if} \quad y_i = y \; , \; f(\mathbf{x}_i) < 0
    \end{cases},
\end{equation}
% \begin{equation}
%     q^{-}_{k,y} = \sum_{(\mathbf{x}_i,y_i) \in D_s} p^{k}_{y}(\mathbf{x}) \quad \textsf{if} \quad y_i = y \; , \; f(\mathbf{x}_i) < 0.
% \end{equation}
namely $q^+$ corresponds to overpredictions $f(\mathbf{x}_i) > 0$ and $q^-$ corresponds to underpredictions $f(\mathbf{x}_i) < 0$. The above assumes equality in the number of data instances per category -- in practice we introduce a per-category scale to account for imbalance. We visually encode $q^{+}$ and $q^{-}$ through a stacked and diverging bar plot, where each attribute is mapped to a row, $q^{+}$ is mapped to the right side of the baseline, and $q^{-}$ is mapped to the left, please see Fig.~\ref{fig:main} for an illustration. For a given attribute, we stack bars based on user-selected categories from the scatterplot. The stacking order is determined based on the sum of the over/under predictions over all attributes, such that categories with a higher sum are positioned closer to the baseline. We also encode this order via a sequential colormap, for a fixed hue of red and blue for over and under predictions, respectively. This view is central in our design, to support precise comparisons between categories and attributes. For example, in Fig.~\ref{fig:main} explainable mispredictions can be quickly determined (\textbf{T2}), e.g. leopards, wolves, and foxes are poorly categorized because the model underpredicts the attribute \textit{strong}, and overpredicts the attribute \textit{brown}.

We populate an additional view that decomposes the data used to compute $q^{+}_{k,y}$ or $q^{-}_{k,y}$ in terms of their false positives (\textbf{T2}), shown in Fig.~\ref{fig:teaser}(D). Upon hovering over a bar, this view is populated such that mispredicted categories are mapped to rows, and each bar encodes the summation factors in $q$ specific to its category. 

\begin{figure}[!t]
	\begin{center}
		\includegraphics[width=.8\linewidth]{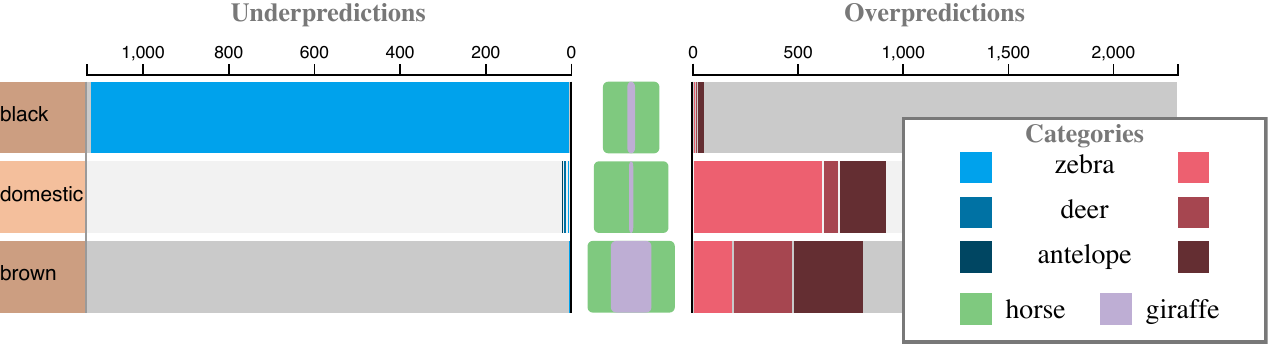}
    \end{center}
	 \caption{Our design for encoding unseen category attributes, alongside attribute errors in seen category, enables visual analysis between seen and unseen categories.}
	 \label{fig:unseen}
\end{figure}

\subsection{Analyzing Unseen Categories}

Only analyzing seen categories tells us little about the model's behavior on unseen categories. We only have access to attribute vectors for unseen categories, and not specific data instances. Yet our visualization is designed around attributes, and thus, we visually encode the relationship between attribute vectors of unseen categories, with attribute errors made on seen categories (\textbf{T3}).

To combine the attribute vectors of unseen categories with the main view, we length-encode the attributes by placing them in the center of the diverging stacked bar chart. Based on the selection made by the user we support 2 different analysis scenarios:

\textbf{Detailed Category Analysis.} If only a single unseen category is selected, each attribute of this category is encoded with a bar in the center, allowing us to relate unseen category attributes with seen category errors. If exactly 2 unseen categories are selected, we enable the user to compare unseen categories, where a category with lower attribute is layered on top of the category with higher attribute -- please see Fig.~\ref{fig:unseen} for such a case.

\textbf{Unseen Category Overview.} If a user selects more than 2 categories, we treat this as a case of analyzing category overviews. We use an ordinal color map, where we map the total number of categories that overlap in a particular attribute to a grey-scale value, shown in Fig.~\ref{fig:teaser}(C) highlights this case.

Additionally, to help the user find visually salient patterns between error in seen category predictions and the unseen category attributes, inspired by LineUp~\cite{gratzl2013lineup} we allow the user to sort by the under/over prediction scores, the sum of the over and under prediction scores, and the sum of attribute values for all selected unseen categories. The different ordering schemes allow the user to prioritize their analysis.%We highlight different strategies in the use cases in Sec.~\ref{sec:casestudy}.

\begin{figure*}[!t]
	\centering
	\begin{subfigure}[b]{0.8\textwidth}
		\includegraphics[width=\linewidth]{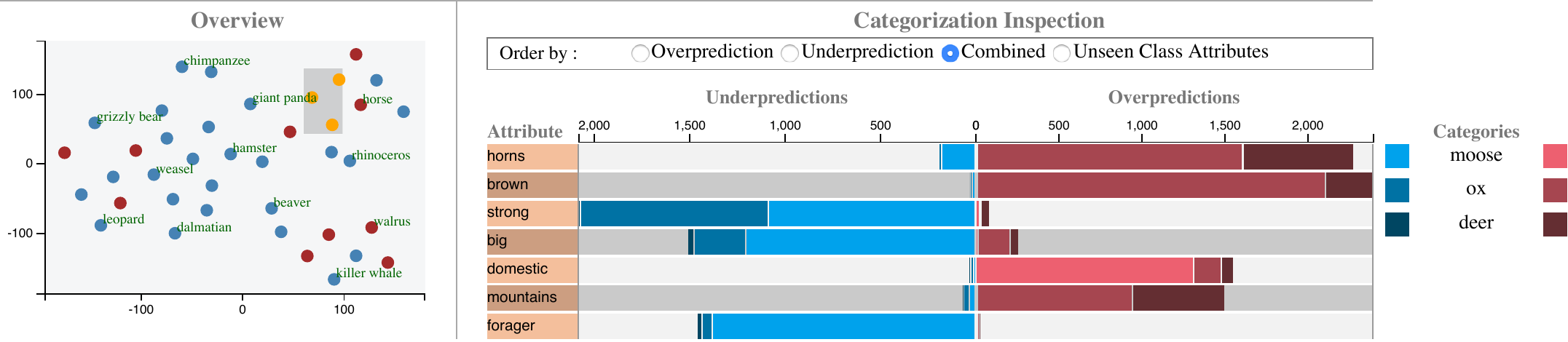}
		\caption{Selection of a small group of seen categories.}
		\label{subfig:cs2-overview}
	\end{subfigure}
	\begin{subfigure}[b]{0.4\textwidth}
	    \centering
		\includegraphics[width=0.84\linewidth]{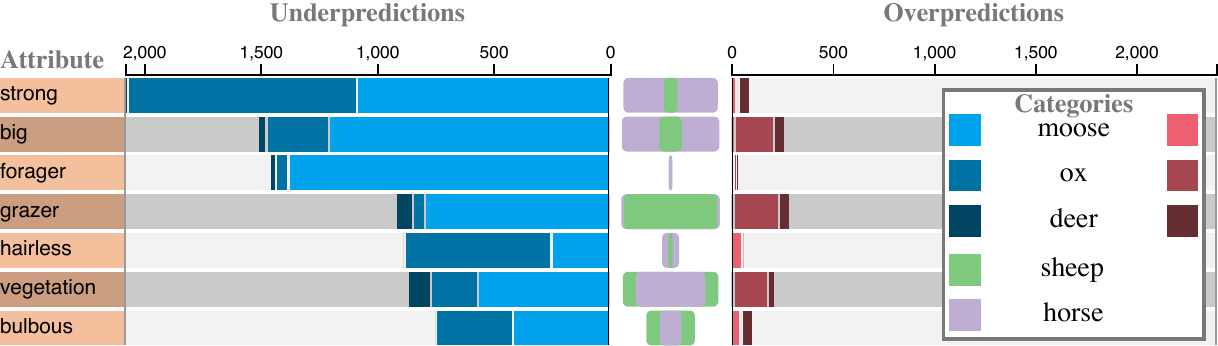}
		\caption{Selecting two unseen classes and ordering by underpredictions.}
		\label{subfig:cs2-under}
	\end{subfigure}
	\begin{subfigure}[b]{0.4\textwidth}
	    \centering
		\includegraphics[width=0.84\linewidth]{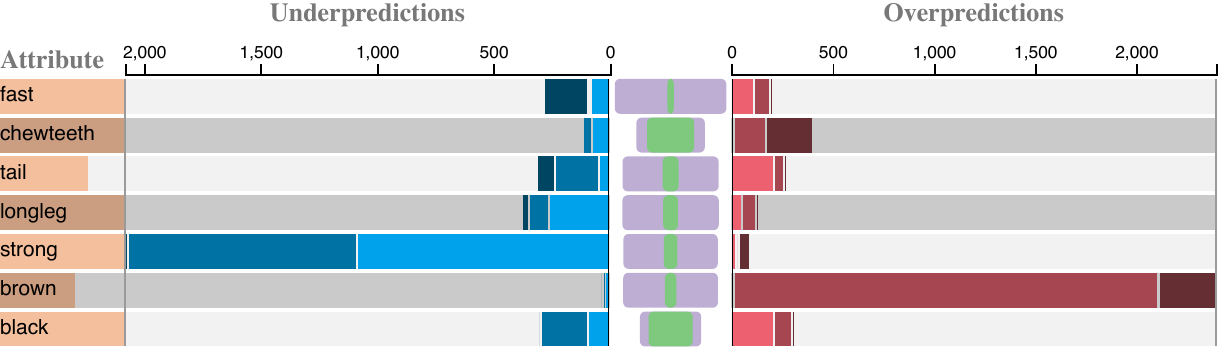}
		\caption{Ordering by unseen class attributes.}
		\label{subfig:cs2-unseen}
	\end{subfigure}
	 \caption{We show a scenario for locally exploring categories. First, the user selects seen categories close together in the scatterplot (a). The user then selects nearby unseen categories, ordering attributes by underpredictions (b). Sorting by summed unseen category values (c), the user can find attributes of different values for unseen categories, and subsequently relate unseen and seen categories in terms of misprediction scores.}
	 \label{fig:cs2}
\end{figure*}

\subsection{Model Steering}
\label{sec:model_steering}
As part of the user's exploration, we support model steering by decreasing the importance of individual attributes, and retraining the model from these weights. Specifically, when a user observes problematic attributes, they may click on any bar of that attribute in the main view, and this will decrease the importance of that attribute (\textbf{T4}). Fig.~\ref{fig:teaser} highlights the results of this process, where in (E) we visually encode the weights by the length of a layered orange-colored bar. Initially all weights are set to 1, and each time the user clicks on a bar we decrease the corresponding weight by 0.1.

To retrain the model we form a diagonal matrix $D \in \mathbb{R}^{a \times a}$, where $D_{ii}$ contains the user's weight for attribute $i$. We then modify our compatibility function by incorporating this matrix, leading to:
\begin{equation}
    s_D(\mathbf{z}_i,\mathbf{z}_j) = \mathbf{z}_i^{\intercal} D \mathbf{z}_j.
\end{equation}
We then substitute $s$ with $s_D$ in Eq.~\ref{eq:maxmargin}, and optimize the resulting loss. The model thus gives less importance to attributes found problematic by the user, and to focus on more reliable attributes. 

\section{Case Studies}

\label{sec:casestudy}
%In this section, we illustrate one of the use cases of our system: global analysis for obtaining a general understanding of seen and unseen categories
% , and local analysis for performing more fine-grained comparisons between categories.

% In this section we illustrate a use case of our system. 
% In our experiments we use the AwA dataset~\cite{lampert2009learning,xian2018zero}. This dataset is comprised of 50 categories corresponding to different types of animals and 85 attributes, where we use ImageNet-trained ResNet features~\cite{he2016deep} of the images as our data input, provided by Xian et al.~\cite{xian2018zero}.
We present a use case of our system, to show how one would analyze the potential behavior of unseen categories.
In our experiments we use the Animals with Attributes dataset~\cite{lampert2009learning,xian2018zero}. This dataset is comprised of 50 animal-based categories and 85 attributes, where we use ImageNet-trained ResNet features~\cite{he2016deep} of the images as our data input, provided by Xian et al.~\cite{xian2018zero}. On average, there are approximately $750$ images per category in the dataset. Please refer to the supplementary material for further details regarding the model, training, and testing

\textbf{Step 1.} The user first selects a subset of categories in the scatterplot to obtain an overview of attribute mispredictions, c.f. Fig.~\ref{subfig:cs2-overview}. They focus their attention on a small cluster of categories -- \textsf{moose}, \textsf{ox}, and \textsf{deer} -- all of which share semantically-similar attributes.

\textbf{Step 2.} The user then finds unseen categories in the scatterplot that are close to the seen categories, and clicks on \textsf{sheep} and \textsf{horse}. The unseen category attribute view is then populated in the center of the baseline, where the user first orders the attributes by underpredictions, shown in Fig.~\ref{subfig:cs2-under}. Here we can observe that sheep are weaker and smaller than horses, and for these attributes we observe high underpredictions over related seen categories, thus there is the potential to incorrectly classify a horse as a sheep by underpredicting \textit{strong} and \textit{big}. Motivated by this observation, the user then prefers to see other attributes where horses and sheep differ, and thus reorders by the sum of the unseen attribute values as shown in Fig.~\ref{subfig:cs2-unseen}. Within this new ordering, the user discovers a set of attributes that have similar characteristics regarding horse and sheep. In particular, they find the attribute \emph{brown} is highly overpredicted, and that horses are generally considered more brown than sheep. Thus, the model is likely to mispredict images of sheep as horses on the basis of attribute \textit{brown}. Furthermore, the \textit{tail} attribute is shown to be equally likely to overpredict or underpredict, indicating a rather unreliable attribute, which could potentially impact the categorization of unseen classes.

\textbf{Step 3.} After identifying unreliable attributes like \textit{brown} and \textit{tail} the user next steers the model by decreasing their weights, as shown in Fig.~\ref{subfig:cs2-unseen}. By retraining the model, we find that the user is able to improve the accuracy of predicting category \textsf{sheep} from 47.6\% to 81.1\%, while also improving the overall performance of the model from 53.2\% to 55.1\%.

In general, we find this localized analysis applies quite well to other unseen categories. We have performed similar model steering experiments centered on other categories, and found encouraging results -- please see the supplemental material for further results. One potential downside to local analysis is that an improvement in one category might result in a decreased accuracy for another category, leading to potential bias. Global analysis of unseen categories -- as depicted in Fig.~\ref{fig:teaser} -- can help counter such bias, thus a mixture of these two analyses is most likely to be useful in practice.

\section{Discussion}

We think that our approach is an important step in using visual analytics to help understand zero-shot learning, and there are several directions we would like to explore as part of future work. We have, thus far, only considered our approach for the Animals with Attributes dataset, and so we intend to use our interface for other ZSL datasets. Though our approach should scale well to datasets such as CUB~\cite{wah2011caltech}, which is comprised of 200 categories and 312 attributes, larger datasets such as ImageNet, comprised of 1,000 categories, will necessitate different visual encodings and interactions. Our diverging, stacked bar design is most useful for comparing tens of categories across several attributes at a time, thus alternative designs, ones that carefully aggregate both category and attribute-based information, will be developed to support the analysis of larger-scale datasets.  We also found it challenging to properly decrease attribute weights through our model steering, as setting attribute weights to be too small can adversely impact performance, though we generally found positive results by decreasing weights to the range of [0.5,0.7]. The problem of translating human judgement of model errors to model weights~\cite{evans2018predicting}, however, is quite challenging, and we believe outside of the scope of our work.

\bibliographystyle{abbrv-doi}

\bibliography{zsl}
\end{document}